\documentclass{article}

\usepackage[final]{neurips_2021}

\usepackage{graphicx}
\graphicspath{ {images/} }

\usepackage[utf8]{inputenc} 
\usepackage[T1]{fontenc}    
\usepackage{hyperref}       
\usepackage{url}            
\usepackage{booktabs}       
\usepackage{amsfonts}       
\usepackage{nicefrac}       
\usepackage{microtype}      
\usepackage{xcolor}         
\usepackage[english]{babel}
\usepackage{float}

\begin{document}

\title{\emph{De novo} PROTAC design using graph-based deep generative models}

%

\author{%
  Divya Nori$^1$, Connor W. Coley$^{1,2}$ \& Rocío Mercado$^2$ \\
  $^1$Department of Electrical Engineering and Computer Science \\
  $^2$Department of Chemical Engineering\\
  Massachusetts Institute of Technology\\
  Cambridge, MA 02139, USA \\
  \texttt{\{divnor80, ccoley, rociomer\}@mit.edu} \\
}

\maketitle

\begin{abstract}
  PROteolysis TArgeting Chimeras (PROTACs) are an emerging therapeutic modality for degrading a protein of interest (POI) by marking it for degradation by the proteasome. Recent developments in artificial intelligence (AI) suggest that deep generative models can assist with the \textit{de novo} design of molecules with desired properties, and their application to PROTAC design remains largely unexplored. We show that a graph-based generative model can be used to propose novel PROTAC-like structures from empty graphs. Our model can be guided towards the generation of large molecules (30--140 heavy atoms) predicted to degrade a POI through policy-gradient reinforcement learning (RL). Rewards during RL are applied using a boosted tree surrogate model that predicts a molecule's degradation potential for each POI. Using this approach, we steer the generative model towards compounds with higher likelihoods of predicted degradation activity. Despite being trained on sparse public data, the generative model proposes molecules with substructures found in known degraders. After fine-tuning, predicted activity against a challenging POI increases from 50\% to >80\% with near-perfect chemical validity for sampled compounds, suggesting this is a promising approach for the optimization of large, PROTAC-like molecules for targeted protein degradation.
\end{abstract}

\section{Introduction}
Of the 1,200 new molecular entities approved by the FDA between 1985-2021, small molecule drugs approved under a ``New Drug Application (NDA)'' comprise $\sim$80\% of them, with the other 20\% being new biological products \citep{FDAapprovals}. Generally speaking, small molecules are designed to impede the function of biologically-relevant target proteins. Small molecule inhibitors interfere with their targets by accessing specific parts of the protein and binding strongly enough to affect their behavior. However, it is estimated that  $\sim$75\% of the human proteome lacks deep binding sites and is thus ``undruggable'' by traditional small molecule inhibitors \citep{toure2016small}. These so-called undruggable targets are nonetheless implicated in a wide range of diseases, including cancer, autoimmune diseases, and cardio-metabolomic diseases, motivating the development of therapeutic modalities beyond small molecule inhibitors.

\begin{figure}[h]
\includegraphics[width=0.45\textwidth]{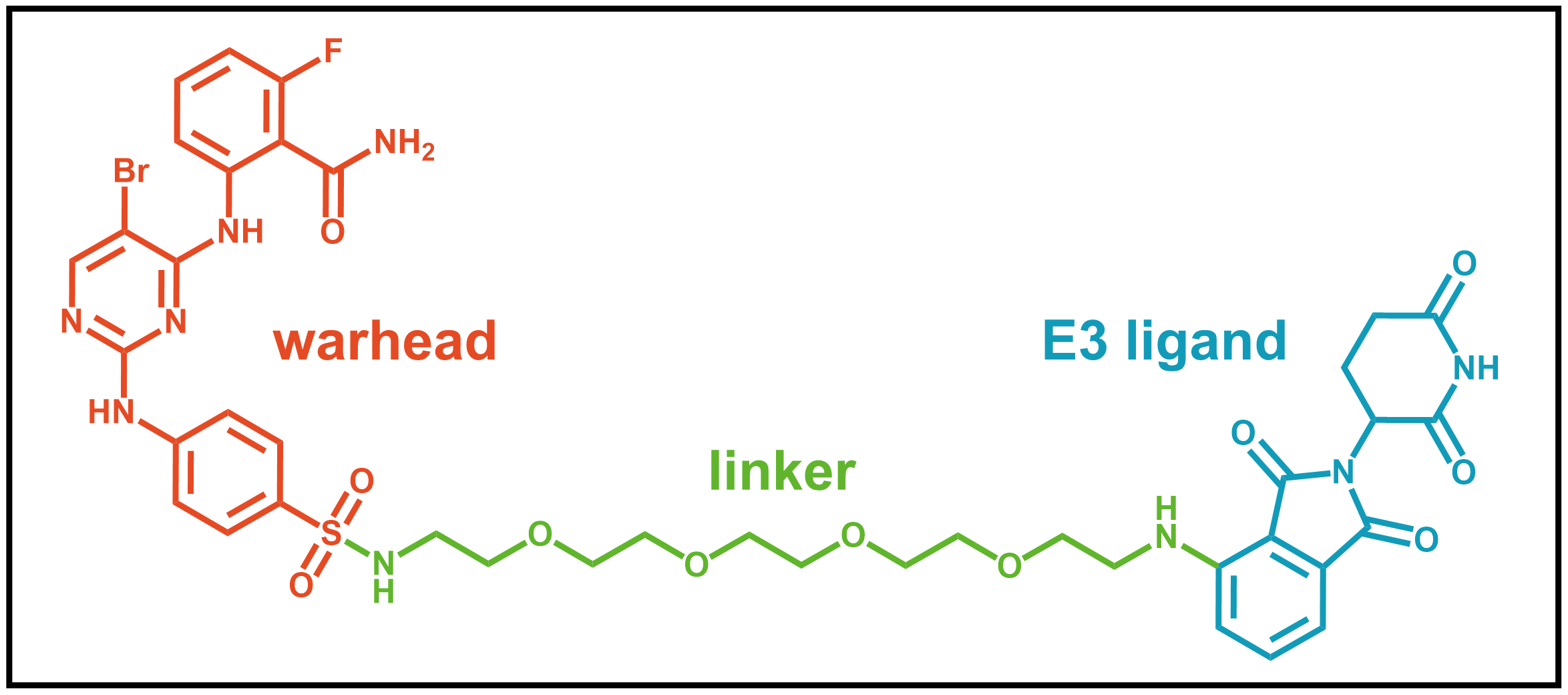}
\centering
\caption{Example PROTAC structure from \texttt{protac-db} (PubChem CID: 155168919), highlighting the three general segments found in PROTACs: the warhead, linker, and E3 ligand.}
\label{fig:protacstructure}
\end{figure}

An example of such an ``undrugabble'' target is the BCL-2 protein, which regulates apoptosis, making it a prime target for cancer drug discovery \citep{bcl2-cancer}. However, in 2021, \citeauthor{bcl2-protac} reported a potent BCL-xL and BCL-2 dual degrader with significantly improved antitumor activity against BCL-xL/2-dependent leukemia cells. This degrader belongs to a class of emerging therapeutic modalities called PRoteolysis TArgeting Chimeras, or PROTACs. As of the end of 2021, there were 15 heterobifunctional PROTACs in clinical development \citep{mullard2021protac}. Generally speaking, though many exceptions exist, the function of synthetic PROTACs is enabled by a three-component structure consisting of two binding domains and an organic linker (Figure \ref{fig:protacstructure}). The two binding domains include a \textit{warhead} designed to bind a protein of interest (POI) and an \textit{E3 ligand} designed to bind an E3 ligase. In the ideal scenario, the linker anchors the two proteins together briefly in a ternary complex, leading to ubiquitination of the POI, which marks it for degradation by the proteasome.

The functionalities of each component are highly interdependent, such that rational design of PROTACs remains challenging. However, recent developments in artificial intelligence (AI) suggest that deep generative models (DGMs) can assist with the \textit{de novo} design of molecules with desired pharmacological profiles \citep{dl-moldesign}. While DGMs have been widely applied to the design of small molecule drugs, DGMs for PROTAC design are typically limited to optimization of the linker. Here, we introduce a graph-based DGM capable of designing PROTAC-like molecules atom-by-atom, starting from empty graphs. We show that the DGM learns to generate PROTAC-like structures containing many substructures found in known degraders, and we show quantitative improvement in a scalar model estimate of degradation activity against a POI. As the model was trained on sparse public data, we did not pursue experimental validation, but instead publish our workflow for PROTAC design open-source. We make the following contributions:
\begin{itemize}
    \item a non-linear, boosted-tree-based model trained on public data for the prediction of protein degradation activity (DC$_{50}$) in PROTAC systems,
    \item application of an existing DGM to distribution-based learning tasks for PROTAC design (30--140 heavy atoms),
    \item application of policy-gradient reinforcement learning (RL) using a multi-objective scoring function to promote the design of structures with predicted protein degradation activity,
    \item a case-study where we apply the above three points to the \textit{in silico} design of novel PROTAC-like structures for IRAK3 degradation.
\end{itemize}

\section{Related work}
\label{gen_inst}
Recent developments in deep learning have led to the emergence of DGMs for \textit{de novo} molecular design \citep{chen2018rise, jimenez2021artificial, meyers2021novo}. Some of the most successful are variations upon RL-based DGMs for the design of drug-like small molecules \citep{rnn, blaschke2020reinvent, gao2022sample}. Policy-gradient RL has been shown to be successful in goal-directed drug design tasks which warrant the prioritization of certain properties, and has been successfully applied to fine-tune DGMs towards the design of molecules with a specific size, octanol-water partition coefficient (logP), or predicted pharmacological activity in a multi-objective fashion \citep{blaschke2020reinvent, rl_graphinvent}.

Previous work on DGMs for PROTAC design has focused on the conditional design of the linker starting from a desired PROTAC substructure. For example, a platform called LinkINVENT used RL to generate favorable connecting components between a pre-specified warhead and E3 ligand \citep{linkinvent}. A graph-based DGM has also been trained to propose 3D linker structures conditioned on partial PROTAC structures \citep{delinker}. More recently, a method called PROTAC-RL was developed which combines a transformer architecture and memory-assisted RL to linker design given an E3 ligand and warhead \citep{protacrl}.

\section{Methods}
\label{headings}

We first discuss the design of a surrogate model for degradation performance, to serve as a PROTAC scoring function. We then describe the graph-based molecular DGM, followed by the integration of the surrogate model into an RL framework and its application towards IRAK3 degrader design.

\subsubsection{Data pre-processing and feature preparation}
\label{sec:embedding}

Data was retrieved from the open-source PROTAC database (\texttt{protac-db}), which has  compiled various experimental measurements from the literature.\citep{protacdbwebsite, protacdbpaper} A single data point includes the PROTAC's SMILES representation; the cell type, E3 ligase, and POI targeted in the experiment; and the DC$_{50}$ value. DC$_{50}$ value gives the concentration of PROTAC needed to degrade 50\% of a POI in a given cell type with a specific E3 ligase. This is a measure of protein degradation activity, where a lower DC$_{50}$ value indicates a more potent PROTAC. The full dataset contains 3,994 datapoints, with 3,270 unique PROTACs represented. Datapoints containing duplicate PROTAC structures are from experiments conducted with the same molecule under different conditions.

To prepare the data, all rows with no explicit DC$_{50}$ value were dropped, resulting in 638 data points. The E3 ligase was represented using a one-hot representation of seven main classes (CRBN, VHL, IAP, MDM2, DCAF, AhR, or RNF), where the most common E3 ligase was cereblon (CRBN). Cell type was one-hot encoded into 148 classes. A comprehensive set of 88 unique  sequences was used to define a vocabulary of bi-grams and tri-grams where each token is an amino acid. The size of the vocabulary was 7,841 words, and each word was used as a feature. PROTACs were represented using 1024-bit Morgan molecular fingerprints, where the fingerprint length was selected from hyperparameter optimization. Embeddings were concatenated into a final embedding with 9,077 features per data point. The continuous response variable was transformed into a categorical variable via the following cut-offs to achieve a balanced class split (appendix Figure \ref{fig:dc50-hist}): DC$_{50}$ $\geq$ 100 nM $\rightarrow$ no to low activity (0); DC$_{50}$ $<$ 100 nM $\rightarrow$ high activity (1). 

\subsection{Surrogate model for protein degradation activity}
\label{sec:surrogate}
To evaluate the quality of PROTAC structures, we developed a surrogate model to predict DC$_{50}$. The model takes as input the aforementioned embedding (section \ref{sec:embedding}). The output is a binary label representing the activity level: 0 (low activity/high DC$_{50}$) or 1 (high activity/low DC$_{50}$). Data was divided into train/test splits using a semi-random 70/30 split, accounting for data-leakage by avoiding having the same PROTAC structure in both splits. Data points where a PROTAC was in the training set were moved out of the test set, leading to 689 training points and 144 points in the hold-out test set. We trained a boosted tree-based model using Light Gradient Boosted Machine (LightGBM version 3.2.1) \citep{lightgbm}. Hyperparameters were selected with Optuna version 2.10.0, using an F1-score objective function and five-fold cross validation \citep{optuna}. Varied parameters included bagging fraction, bagging frequency, learning rate, number of leaves, and feature fraction.

\subsection{Graph-based generative model}
We tackle the challenge of optimizing all three components simultaneously. We do this using GraphINVENT \citep{graphinvent, rl_graphinvent}, a graph-based autoregressive DGM which uses RL for molecular optimization, as it was previously demonstrated to be successful in the generation of large natural products of similar size to PROTACs \citep{mercado2021exploring}. To build the DGM, we followed a three-step workflow: \textit{preprocessing}, \textit{training}, and \textit{generation}. Preprocessing involves creating step-by-step graph reconstruction processes for each molecule in the training set. This step-wise information is then used to train the DGM. As \texttt{protac-db} contains <30 molecules with phosphorous and iodine, these were removed from the dataset for computational efficiency (less padding). The final training set consisted of 4,120 molecules with atom types \{C, N, O, F, S, Cl, and Br\}, formal charges \{-1, 0, and 1\}, and a maximum heavy atom count of 139. This set was used to pre-train the DGM for 200 epochs with a batch size of 50.

\begin{figure} [h]
\begin{center}
\includegraphics[width=13cm]{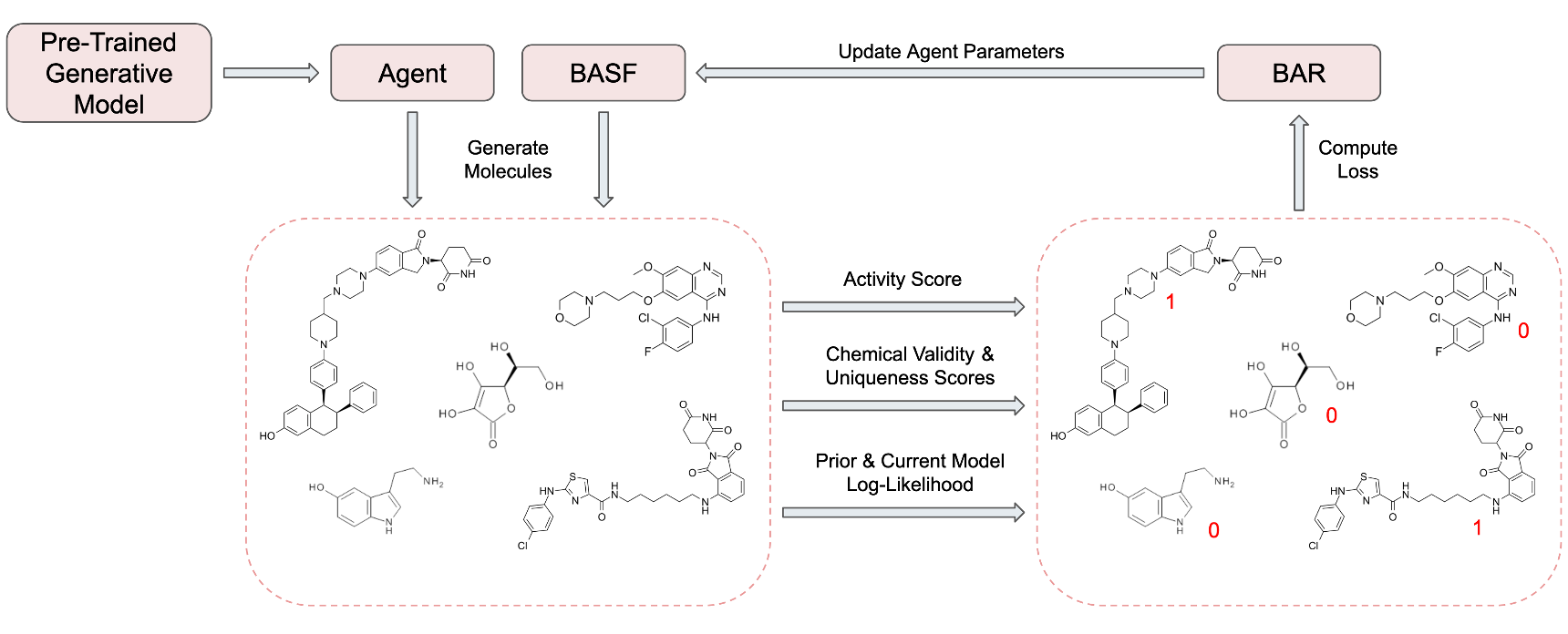}
\caption{Reinforcement learning loop used to fine-tune the pre-trained deep generative model. BASF stands for ``Best Agent So Far'' and BAR represents the  ``Best Agent Reminder'' loss.}
\label{fig:rl-loop}
\end{center}
\end{figure}

\subsection{RL framework for fine-tuning}
We adopt the the memory-aware RL loss used in GraphINVENT \citep{rl_graphinvent} to steer our generative model towards PROTACs more likely to be active using a surrogate model for degradation activity. The framework, illustrated in Figure \ref{fig:rl-loop}, begins with the pre-trained DGM, which is used to initialize the Agent as well as the Best Agent So Far (BASF). Both agents generate a set of molecules, saving the actions. Sampled molecules from each agent are scored by their predicted degradation activity, modified by a term that prevents rewarding duplicate, chemically invalid, heavy atom count <30\footnote{Fewest heavy atoms per molecule in \texttt{protac-db} = 30.}, or improperly terminated molecules. The BASF, prior, and current agent log-likelihoods are used to compute the loss and update the model through gradient descent as described by \citep{rl_graphinvent}. This framework was used to fine-tune the model from the best pre-training epoch, $e^*$, for 200 additional steps. The pre-trained model was fine-tuned 10 times to collect statistics. The agent and BASF each use a generation batch size of six, where sampled molecules are evaluated at each RL step using the above scoring mechanism. For training the agent, the Adam optimizer was used (initial learning rate of $10^{-6}$) with the OneCycle learning rate scheduler, no weight decay, and  $\alpha = 0.5$. 

10K molecules were sampled from the converged agent at 200 fine-tuning steps. The molecules were evaluated for 1) degradation activity using the surrogate model (section \ref{sec:surrogate}), and 2) chemical diversity using Murcko scaffold decomposition. These molecules were compared against the 2.5K molecules sampled from the model at $e^*$ before RL fine-tuning. 

\subsection{Case study: optimizing for IRAK3 degradation}
We selected interleukin-1 receptor-associated kinase 3 (IRAK3) from \texttt{protac-db} for fine-tuning the DGM. IRAK3 has been implicated in oncological signaling, and its inhibition induces T-cell proliferation for reduced tumor burden. IRAK3 contains an ``undruggable'' ATP binding site which has been the target of PROTAC development efforts \citep{irak3}. This is a challenging case study where the model has to learn to generate new PROTACs for IRAK3 degradation having seen one example of IRAK3 degraders previously. 

\section{Results}
\label{headings}

\subsection{Activity scoring model metrics}

\begin{figure} [h]
\begin{center}
\includegraphics[width=14cm]{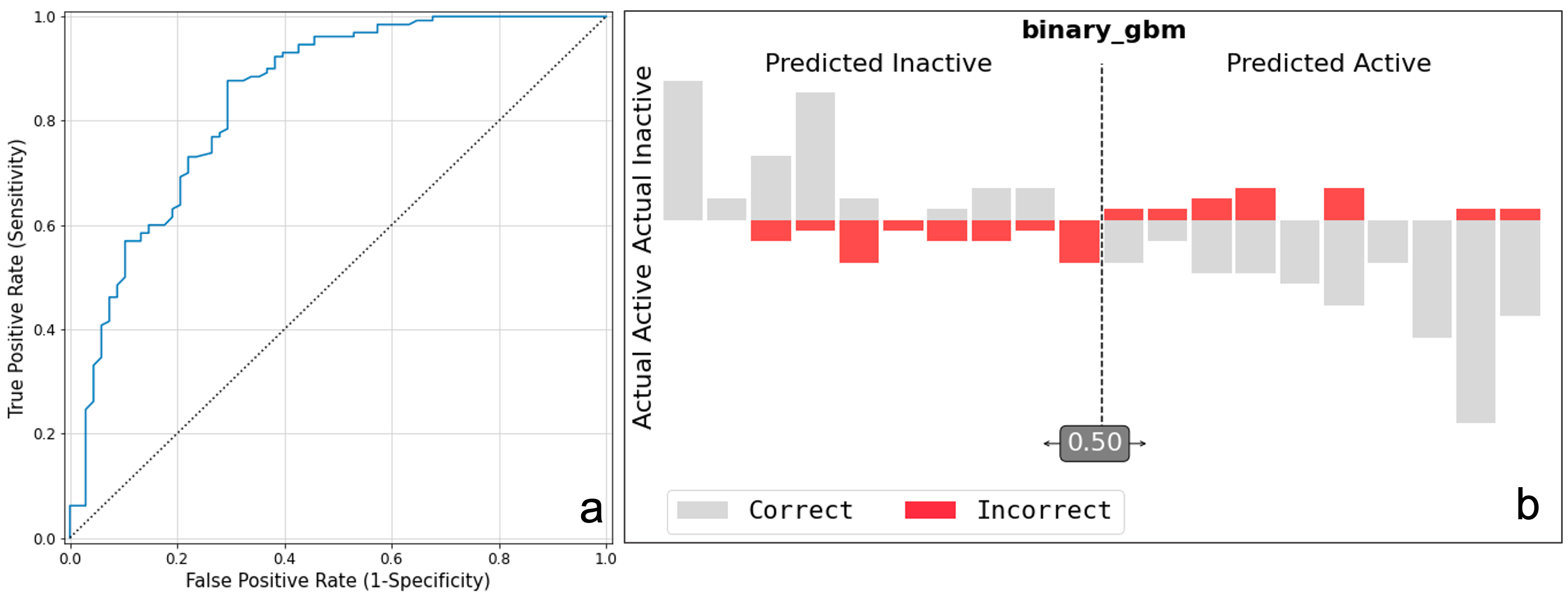}
\caption{(a) Receiver-operating characteristic for surrogate model protein degradation activity as a binary classification task, demonstrating an AUROC of 0.877 on the test set. (b) Analysis of test set prediction accuracy in terms of binned predicted activity score.}
\label{fig:surrogate-model}
\end{center}
\end{figure}

We show here the results of the tree-based activity scoring model on the held-out test set. The model returns a score between 0 and 1 for each input molecule where higher scores suggest greater PROTAC activity. As shown in Figure \ref{fig:surrogate-model}, the final model's test AUC is 0.87, and for molecules with scores $\leq0.3$ or $\geq0.7$, the predicted activity is correct $87.6\%$ of the time. For molecules that score between $0.3$ and $0.7$, the probability of incorrect classification is $59.0\%$. Therefore, when using this model for DGM fine-tuning, molecules that score within this range are not rewarded. However, $90\%$ of molecules in the test set scored outside of this range, indicating the model is relatively confident in those predictions.

\subsection{Fine-tuning DGM to generate active PROTACs}
\begin{figure} [h]
\begin{center}
\includegraphics[width=13cm]{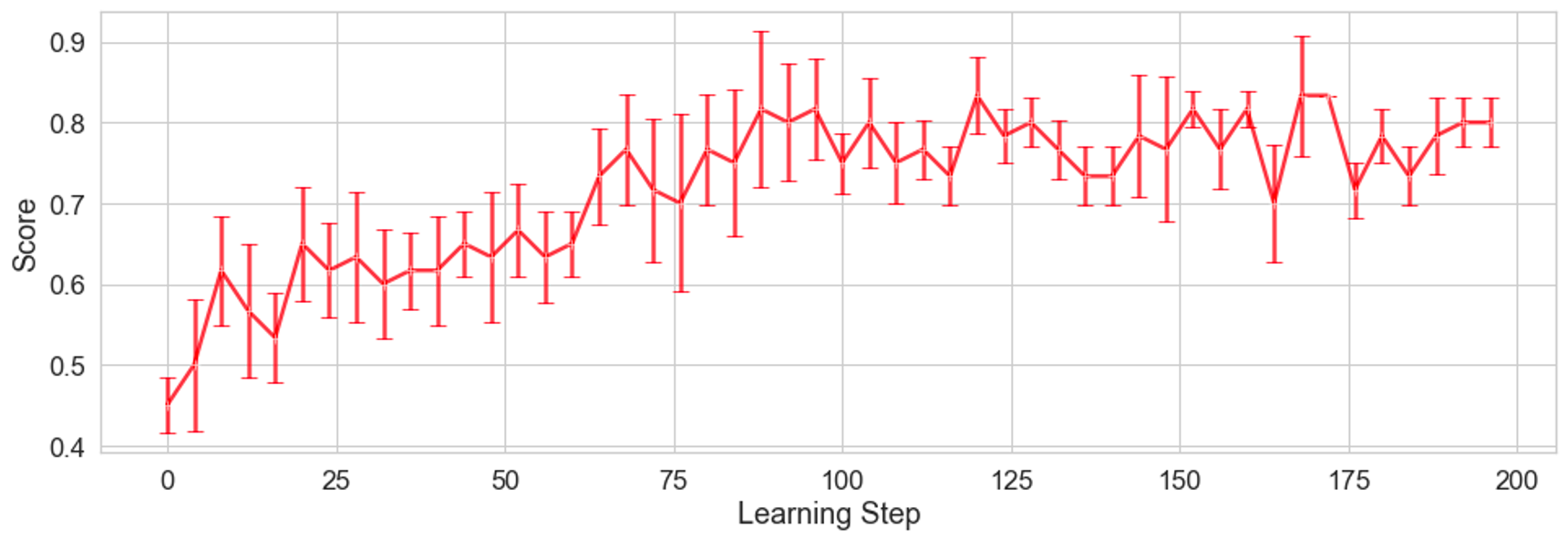}
\caption{Activity score progression across learning steps during RL training. Error bars are the standard deviation from the results of 10 distinct runs.}
\label{fig:score-during-RL-training}
\end{center}
\end{figure}

The DGM was fine-tuned using RL from the pre-trained model state at the epoch, $e^*$, that maximized the fraction of valid molecules in the range of epochs 150--200, as models in this range provide a good compromise between the validity of sampled structures and  validation loss (appendix Figure \ref{fig:validity}). Starting from the pre-trained model at $e^*=192$, we observe that the score of the sampled molecules, interpretable as a binary activity measure, steadily increases for the first 100 RL steps before converging (Figure \ref{fig:score-during-RL-training}), suggesting that the model is learning to generate molecules that score highly for IRAK3 degradation. 

\begin{figure}[h]
\begin{center}
\includegraphics[width=12cm, trim={0 0 0 0}, clip]{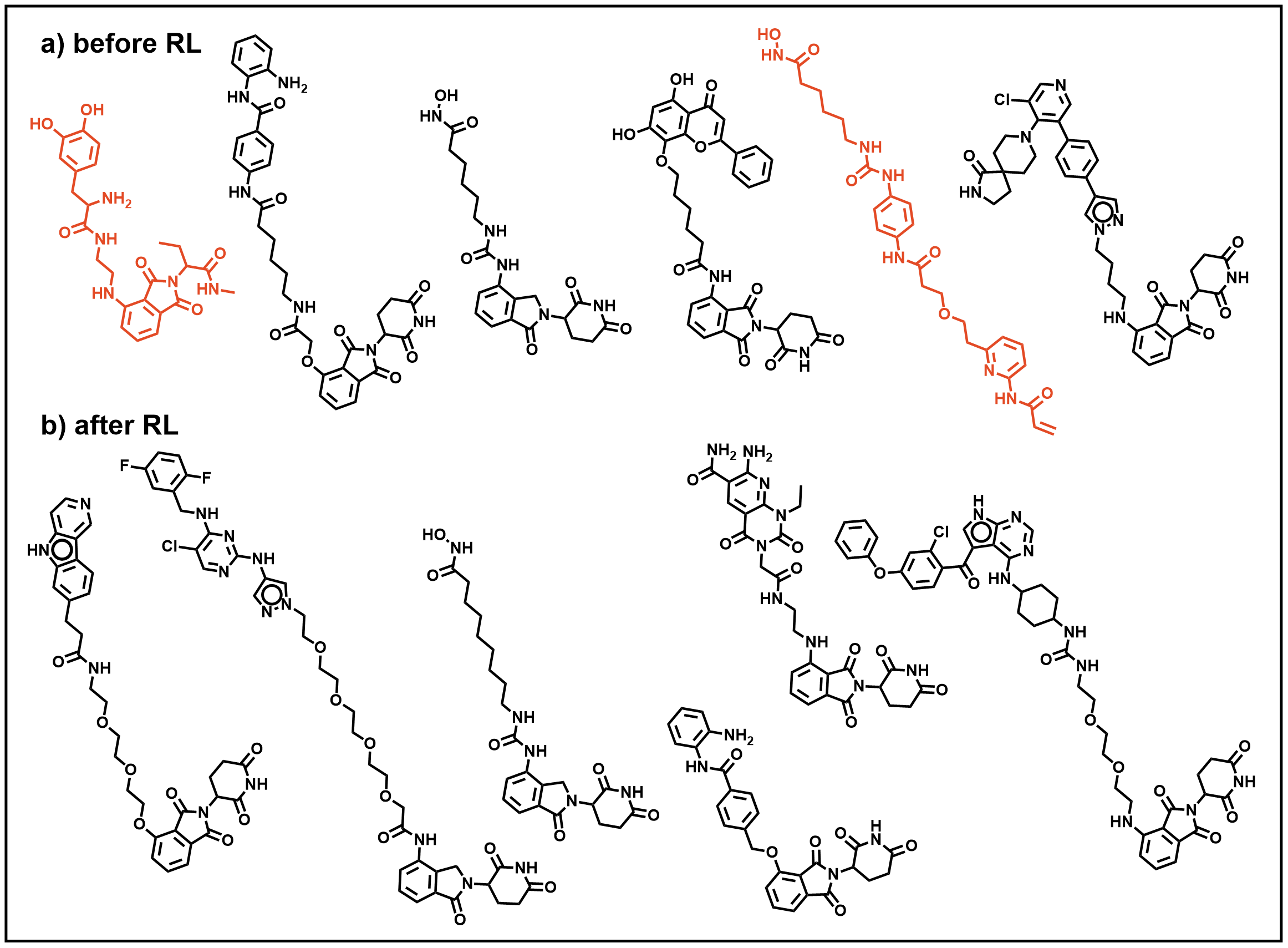}
\caption{Generated molecules randomly sampled from the DGM (a) before and (b) after reinforcement learning (RL). The black molecules are predicted to be active for IRAK3 degradation by the surrogate model, whereas the red molecules  are predicted to be inactive. None of the generated molecules appear in the training set.}
\label{fig:generated-molecules}
\end{center}
\end{figure}

To compare the quality of molecules being generated at these two different states, 2.5K were sampled from the pre-trained model at $e^*$, and 10K molecules were sampled from the fine-tuned model after 200 RL steps. Visualizations of randomly selected molecules from the two sampled sets are shown in Figure \ref{fig:generated-molecules}. What we observe is that at before applying RL, the DGM has already learned to generate barbell-like molecules, characteristic of structures in \texttt{protac-db}. Many of the sampled molecules have well-developed E3 ligands and warheads, thought only about half ($50.8\%$) are predicted to be active degraders. However, after RL fine-tuning, we observe that $83.8\%$ of sampled molecules demonstrate predicted IRAK3 degradation activity, and nearly 100\% chemical validity (appendix Figure \ref{fig:validity}). 

\subsection{Evaluation of PROTACs generated by the best model}
We evaluated the scaffold diversity of the 10K molecules generated from the DGM  after RL. The average number of heavy atoms in this sampled set was 56.46, compared to 55.09 for \texttt{protac-db}. Additionally, both before and after fine-tuning, 0\% of the sampled molecules were regenerated from the training set; in other words, they were all novel structures, likely due to the large number of nodes in the PROTAC graphs and the large action space during generation.

Diversity of generated molecules was evaluated using molecular scaffolds, a concept applied in medicinal chemistry to represent core substructures in bioactive compounds. Specifically, we identified the Murcko scaffolds present in the final 10K set of generated molecules and in the $\sim$5K molecules in \texttt{protac-db}, then computed the intersection of these two sets. Murcko scaffolds were identified using RDKit \citep{rdkit}. In the 10K final generated molecules, we identified 537 unique Murcko scaffolds, whereas in \texttt{protac-db} we identified 2,907 unique Murcko scaffolds. This is not surprising as during fine-tuning we are narrowing in on a smaller chemical space. See appendix \ref{app:murcko} for examples of the most common scaffolds.

\begin{figure}[h]
\begin{center}
\includegraphics[width=12cm]{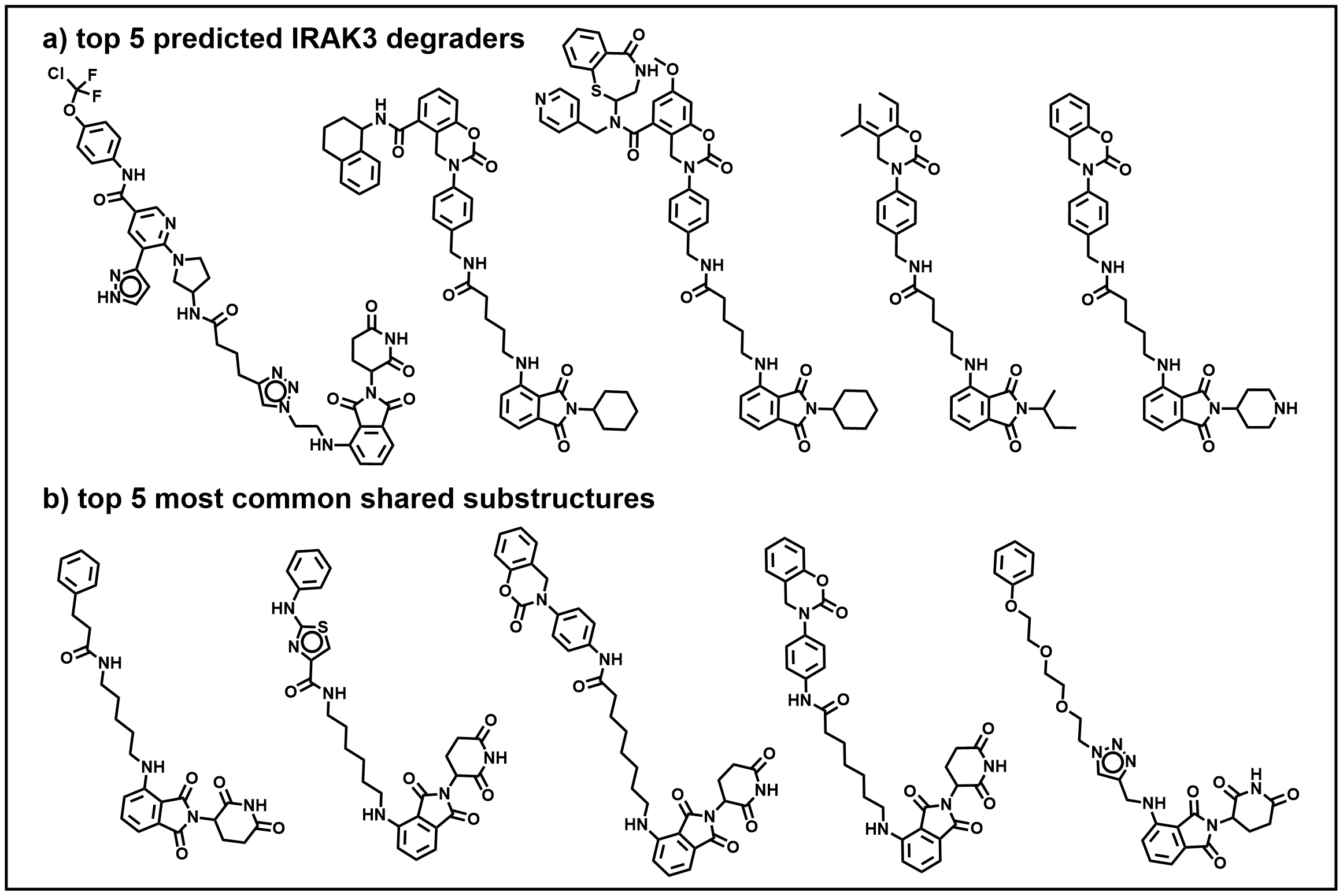}
\caption{(a) Top 5 predicted IRAK3 degraders sampled from the DGM after reinforcement learning. (b) Top 5 most common substructures shared by the top 100 generated molecules and the known actives in \texttt{protac-db}.}
\label{fig:scaffolds}
\end{center}
\end{figure}

In Figure \ref{fig:scaffolds}b, we highlight the most common substructures shared by the top 100 generated and \texttt{protac-db} molecules. We see that the model has learned to reuse several substructures repeatedly. We see similar trends when analyzing the top 5 most common substructures 
from the top 100 predicted IRAK3 degraders (out of the 10K generated set; appendix Figure \ref{fig:common-scaffolds}). Interestingly, the most common substructure found in the top 100 predicted degraders is phthalimidinoglutarimide (PubChem CID: 91585), a known IRAK degrader and CRBN binder \citep{bekes2022protac}. The second most common corresponds to a known degrader for the tau protein (PubChem CID: 137408522). Noticably, all top 5 most common substructures contain the phthalimidinoglutarimide motif (a substructure of thalidomide and lenalidomide), indicating that our model is satisfactorily learning the structure of CRBN ligands. This is not surprising as there is less variation amongst E3 ligands than warheads in \texttt{protac-db}.

\section{Discussion}
\label{headings}
\subsection{A look at the potential new degraders}
Above, we show that a graph-based DGM can learn the structure of potentially new degraders via reinforcement learning such that 82\% of the final sampled molecules are predicted to be potent degraders with DC$_{50}$~<~100~nM when optimized against an example task from \texttt{protac-db}. Many of the sampled molecules contain scaffolds present in known degraders. Although the DGM receives no information during training on which component is the warhead, the linker, or the E3 ligand, it learns to generate promising new PROTAC-like structures based on the final reward for the entire output molecule. It also does this without any initial seed substructure.

While previous work has primarily applied generative models to design small molecules containing $\leq 30$ heavy atoms, here we show that our DGM can be used to generate molecules containing up to 139 heavy atoms with nearly 100\% chemical validity (appendix Figure \ref{fig:validity}). This indicates that the DGM has learned chemical rules and can apply them to build large, complex molecules.

\subsection{Limitations}
The primary avenue for improvement in this work is in increasing the generalizability of the DC$_{50}$ prediction model. In this work we were limited by the sparse nature of the public PROTAC data. Several data points in \texttt{protac-db}, which are mined from the literature, lacked DC$_{50}$ measurements, making those entries unusable. In addition, there is considerable class imbalance between E3 ligases, cell types, and POIs, which affects the ability of the model to fully utilize the E3 ligase and cell type information in the dataset (see appendix \ref{app:ablation} for an ablation study). While some ligases have hundreds of measured data points, others have just a few. Addressing these data and representation limitations could enable the application of more complex architectures for the degradation prediction model, and also improve the DGM which builds on the accuracy of the surrogate model for PROTAC design. Another limitation is that synthesizability considerations have not been factored in to the DGM in this work, which will be relevant when it comes to experimental validation and are thus important to address, along with the aforementioned issues, in future work. 

\subsection{Future work}
There remains much to improve for the creation of an automated PROTAC design pipeline; nonetheless, one main area of improvement is the accuracy of the DC$_{50}$ surrogate model. Better representations couldbe used for E3 ligase and cell type information. For instance, E3 ligase sequence information could be included, and cell type embeddings could be constructed to encode biological similarity between types. Additionally, besides the DC$_{50}$ values listed in \texttt{protac-db} for the response variable, docking studies could be used to estimate binding affinities between the PROTAC, E3 ligase, and POI.\citep{zaidman2020prosettac, bond2021proteolysis} 

While the molecules generated are predicted to be active by the surrogate model, they cannot be deemed true actives from computational predictions alone. Experimental validation \textit{in vitro} is essential for moving forward in the drug development process of any compound designed \textit{in silico}.

\section{Conclusions}
\label{headings}
Here, we have demonstrated how a graph-based DGM can be directed towards the generation of predicted active PROTACs via policy-gradient RL using a memory-aware loss function. It can be used to design PROTAC-like structures with up to 140 heavy atoms. The scoring model used in the RL framework is a boosted tree-based classification model for protein degradation activity; after hyperparameter tuning, the model displays a test AUC of $0.87$, though the generalizability of the model is limited due to the sparse nature of existing database. RL enhances the percentage of predicted active PROTACs generated by the DGM from $53\%$ to $82\%$ in 200 learning steps. Analysis of molecules sampled from the final fine-tuned model shows that while the compounds are 100\% novel, they contain substructures present in known protein degraders. Generally, we have shown that our graph-based model can be used to optimize large therapeutic molecules such as PROTACs. With the availability of better public data, and the development of better physics-based models for ternary structure modeling, machine learning tools can be used to make the PROTAC design process less formidable. We hope this work inspires future research on \textit{de novo} design tools for emerging therapeutic modalities.

\section{Acknowledgements}
\label{headings}
D.N. thanks the MIT UROP office for a summer fellowship. R.M. was financially supported by the Machine Learning for Pharmaceutical Discovery and Synthesis consortium. We thank Samuel Goldman for useful feedback and discussions.

\section{Code availability}
\label{headings}
Code for the surrogate model is available on GitHub at \url{https://github.com/divnori/Protac-Design}. Modified GraphINVENT source code and generated data, including pre-trained models, fine-tuned models, and generated structures, can be found at \url{https://doi.org/10.5281/zenodo.7278277}. \\

\bibliography{sample}
\label{headings}

\section{Appendix}
\label{headings}

\subsection{Feature importance ablation study}
\label{app:ablation}
Feature importance was measured via an ablation study using input feature shuffling on the held-out test set. For a given embedding, the respective column values were shuffled in the input matrix to randomize the values for those variables. The scoring model was then used to predict the DC$_{50}$ class from the modified input. The results are shown in Table \ref{tab:feature-importance}, where ``Importance'' is calculated by subtracting the AUC and F1 score sum from the original (unshuffled) model's AUC and F1 score sum. Input shuffling occurs with slight variation on each pass, so the experiment was conducted three times and used to compute the standard error of the mean (SEM) for each importance value.

\begin{table}[h]
  \centering
  \begin{tabular}{lllll}
    \toprule
    \cmidrule(r){1-2}
    Shuffled Embedding     & AUC     & F1 Score & Importance & SEM\\
    \midrule
     None (Original)  & 0.860 & 0.893  & NA & NA\\
    PROTAC structure & 0.611 & 0.832  & 0.310 & 0.012 \\
    Receptor (POI) & 0.765  & 0.838  & 0.150 & 0.023 \\
    E3 Ligase & 0.855  & 0.893  & 0.005 & 0.000 \\
    Cell Type  & 0.860 & 0.893  & 0.000 & 0.000 \\
    \bottomrule
  \end{tabular}
  \caption{Feature importance in the surrogate model.}
  \label{tab:feature-importance}
\end{table}

From Table \ref{tab:feature-importance}, we observe that the PROTAC molecular fingerprints (1024-bits) are the most important. The receptor n-grams (7,841 bi-gram \& tri-gram features) follow, with the one-hot encoded E3 ligases (7 features) and cell types (148 features) being the least important. This shows that the model learns from both the PROTAC structure and receptor to reliably predict PROTAC degradation activity for the POI. In comparison, E3 ligase has relatively low feature importance, whereas the cell type is seemingly not contributing at all to the model's learning under the current featurization scheme.

\begin{figure} [h]
\begin{center}
\includegraphics[width=7cm]{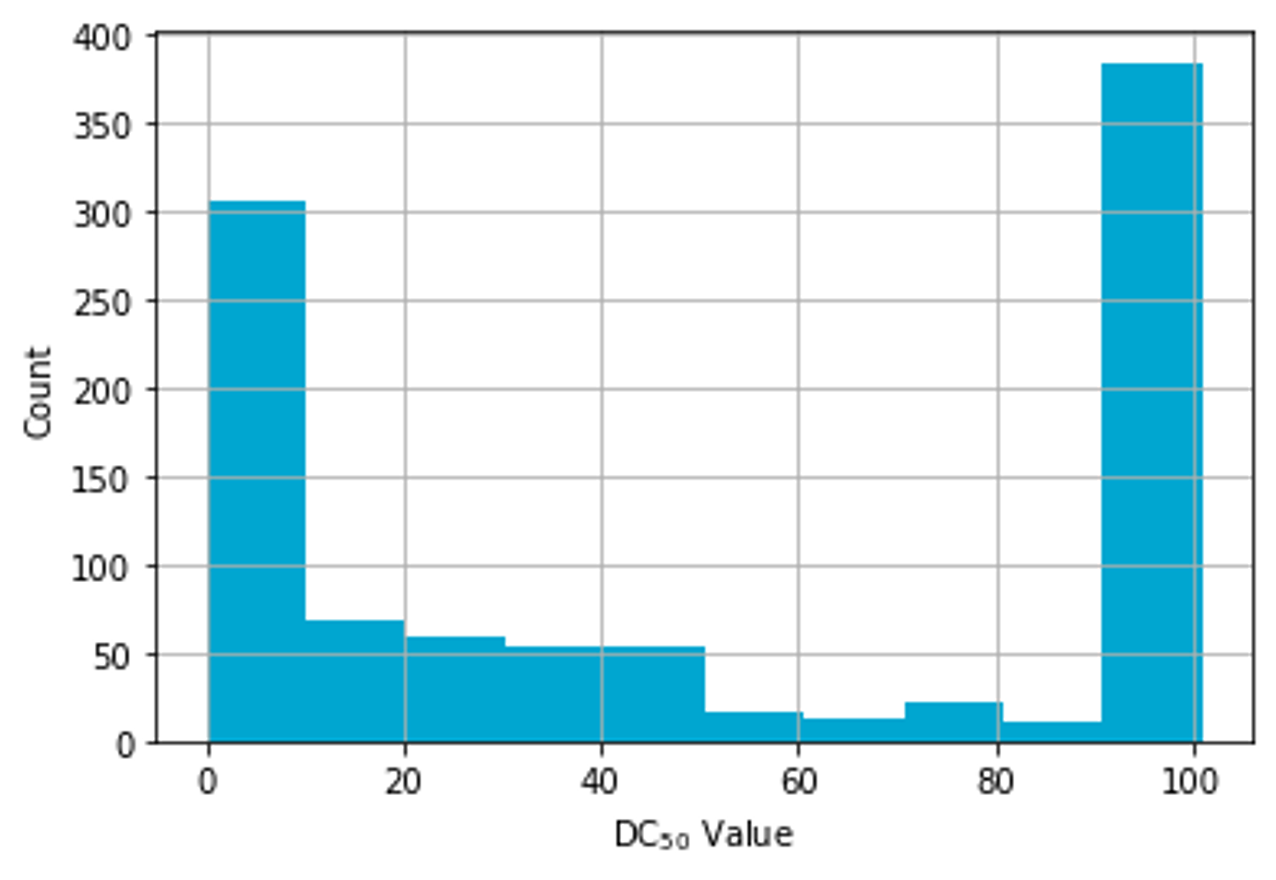}
\caption{DC$_{50}$ histogram for the 638 \texttt{protac-db} datapoints used in this work.}
\label{fig:dc50-hist}
\end{center}
\end{figure}

\subsection{Response variable preparation}
Figure \ref{fig:dc50-hist} shows the histogram which was used to determine the threshold for encoding the response variable, DC$_{50}$, into balanced binary classes. Based on this distribution, a threshold of 100 nM was used to split DC$_{50}$ values into two classes. All DC$_{50}$ values <100 were put into one class (1: ``high activity''), and the remaining values were put into another class (0: ``no to low activity'').

\begin{table}[h]
  \label{sample-table}
  \begin{tabular}{llllll}
    \toprule
    F1 Score     & Bagging Fraction     & Bagging Freq & Learning Rate & Number of Leaves & Feature Frac \\
    \midrule
    0.8316 & 0.8876  & 3  & 0.2299 & 19 & 0.5412  \\
    0.8288 & 0.8153  & 3  & 0.2936 & 21 & 0.4045  \\
    0.8287 & 0.8784  & 3  & 0.2957 & 20 & 0.3990  \\
    0.7126 & 0.6383  & 5  & 0.0346 & 17 & 0.4954  \\
    0.6981 & 0.4260  & 4  & 0.0123 & 24 & 0.4600  \\
    0.6953 & 0.5313  & 6  & 0.1922 & 11 & 0.5114  \\
    \bottomrule
  \end{tabular}
  \caption{Surrogate model parameters varied during hyperparameter optimization.}
  \label{tab:hyperparameters}
\end{table}

\subsection{Hyperparameter optimization}
Table 2 shows the hyperparameters tuned during surrogate model training. The top three rows show the best three hyperparamater combinations, ranked using an F1 objective function. The bottom three rows show the worst three combinations.

\begin{figure} [h]
\begin{center}
\includegraphics[width=13cm]{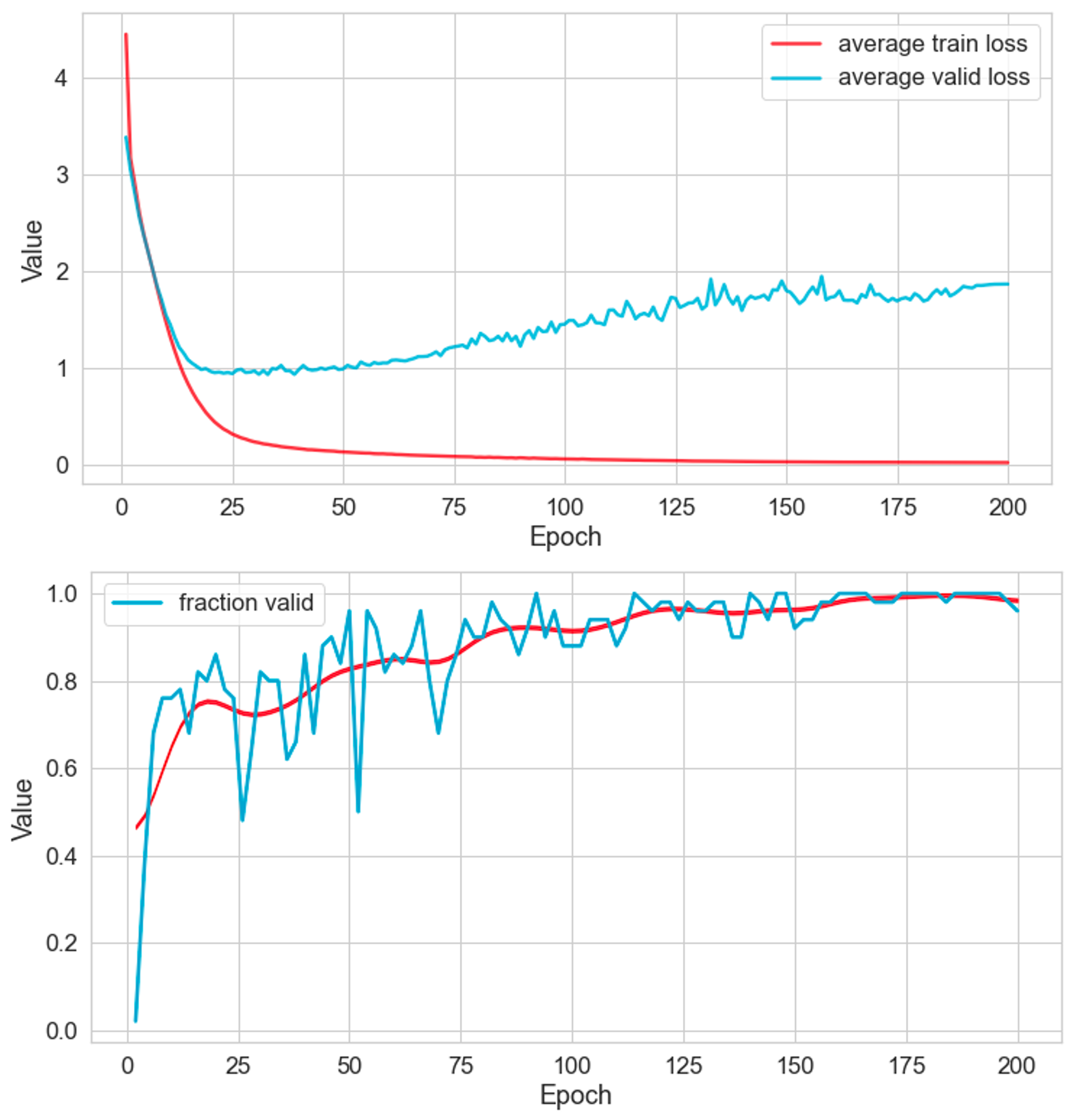}
\caption{Training loss, validation loss, and fraction valid as a function of pre-training epochs.}
\label{fig:validity}
\end{center}
\end{figure}

\subsection{Additional training results}
In Figure \ref{fig:validity}, we show the training/validation loss and fraction of valid molecules sampled from the DGM as a function of training epochs. Initially, as the number of epochs increases, the average validation loss increases. Subsequently, the validation loss plateaus, indicating some amount of overfitting. The fraction of valid molecules reaches 100\% by the final epoch.

\subsection{Most common substructures}
\label{app:murcko}
\begin{figure} [h]
\begin{center}
\includegraphics[width=12cm]{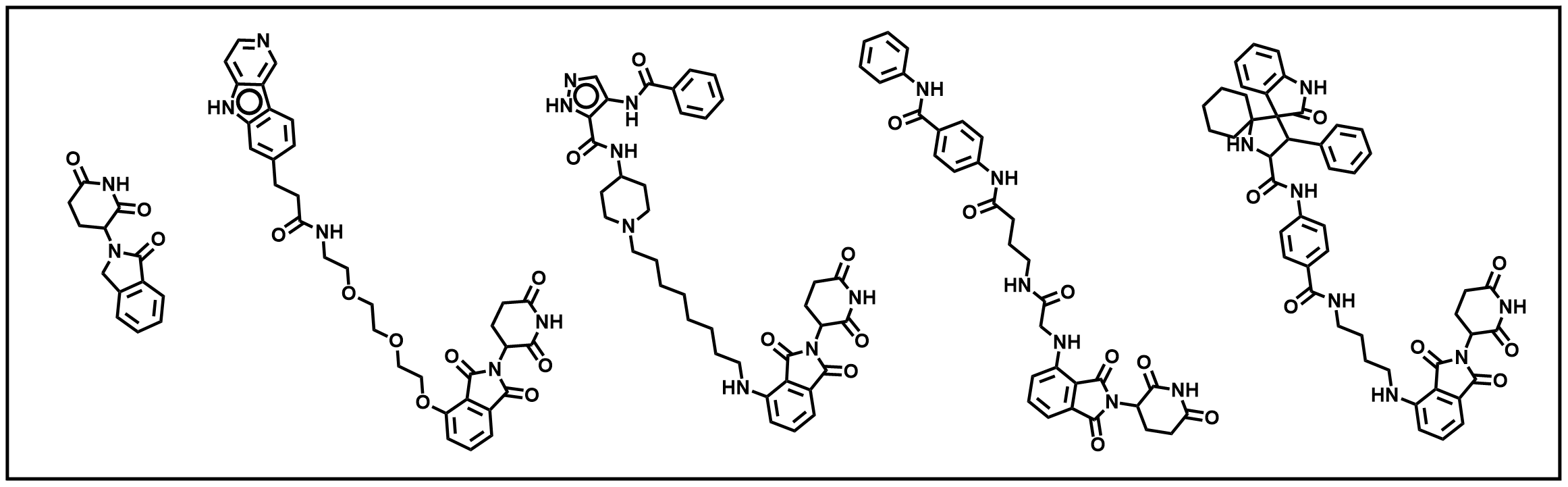}
\caption{Most common substructures appearing in the top 100 predicted degraders for IRAK3, in order of most common (left) to less common (right).}
\label{fig:common-scaffolds}
\end{center}
\end{figure}

In Figure \ref{fig:common-scaffolds} we illustrate the most common substructures present in the sampled molecules predicted to be most active after RL. 
The two left-most molecules in Figure \ref{fig:common-scaffolds} are known protein degraders.

\end{document}